\documentclass[a4paper,twoside,reqno]{bjp}
\usepackage{graphicx}
\usepackage{cite}
\usepackage{amssymb,amsmath,amscd,amsthm}
\usepackage{times}

\usepackage[bookmarks=false]{hyperref}
\hypersetup{%
	colorlinks=true,        % false: boxed links; true: colored links
	linkcolor=blue,          % color of internal links (change box color with linkbordercolor)
	citecolor=blue,         % color of links to bibliography
	urlcolor=blue           % color of external links
}

\usepackage{geometry}
\allowdisplaybreaks

\begin{document}
	
	\title{Study of few $^3$He-induced nuclear fusion reactions using density-dependent double-folding complex potential}
	% Insert the title
	
	\runningheads{N. Mohammad, H. Sultana, Md. R. Islam, Md. A. Khan}{Study of few $^3$He-induced non-resonant nuclear fusion reactions ....potential} 
	% In case the authors are more then three, put the name of the first author followed by the Latin `et al.'
	
	\begin{start}{%
			\author{Nur Mohammad}{1},
			% The second argument connects author(s) with addresses
			\author{Md. Rabiul Islam}{1},
			% The second argument connects author(s) with addresses
			\author{Hajera Sultana}{2},
			% The second argument connects author(s) with addresses
			\author{Md. Abdul Khan}{2*}
			
			\address{Department of Physics, Raiganj University, Raiganj, Uttar Dinajpur, WB-733134, India}{1} 
			\address{Department of Physics, Aliah University, IIA/27, Newtown, Kolkata-700160, India; *Corresponding author E-mail: drakhan.rsm.phys@gmail.com; drakhan.phys@aliah.ac.in}{2} 
%			\address{*Corresponding author E-mail: drakhan.rsm.phys@gmail.com; drakhan.phys@aliah.ac.in}	

			\received{Day Month Year (Insert date of submission)}
			% Insert date of submission
				}

\begin{Abstract}
Nuclear fusion reactions at sub-barrier energies play crucial roles in many aspects of primordial nucleosynthesis in stellar objects. One of the primary aspects that plays a pivotal role in understanding the relationship between stellar evolution and nuclear reaction dynamics is the energy dependence of astronuclear observables, such as the fusion cross-section $\sigma$. This paper presents the results of a few  $^3$He-induced nuclear fusion reactions-$^{3}$He($^3$He,2p)$^{4}$He, $^{6}$Li($^3$He,d)$^{7}$Be and $^{10}$B($^3$He,n)$^{12}$N which are investigated adopting the single-step selective resonant tunnelling model (SRTM). As an improvement over earlier works, the authors have used a microscopically derived density-dependent double-folding potential model, invoking the M3Y-Reid NN interactions, for the numerical computation of the astrophysical S-factor, $S(E)$, and the fusion cross-section, $\sigma$. The results of the calculations have been compared with those found in the literature. The results obtained in the present studies agree fairly with the experimentally observed results found in the literature. 
		\end{Abstract}
		
		\begin{KEY}
			Fusion cross-section (FSC), M3Y-Potential, primordial nucleosynthesis, double-folding potential (DFP)
		\end{KEY}
	\end{start}
	 PACS Number(s): 25.70.Gh; 25.60.Pj

	%%%%%%%%%%%%%%%%%%%%%%%%%%%

\section{Introduction:}
Astrophysics is one of the foremost subjects studied on the origin and evolution of chemical elements\cite{bonnell-2002,carpenter-2000,caselli-2012}. Nuclear astrophysics can be viewed as a convergence of nuclear physics and astrophysics. This emerging branch of physics plays a crucial role in the fruitful description of several physical phenomena, including energy production in stars, nucleosynthesis, production of superheavy elements, etc. \cite{black-2014,bethe-1939}. The explanation of the relative abundance of light nuclides ($^2$H, $^3$He, $^4$He, $^7$Li) and the theoretical description of the nucleosynthesis are the prime achievements of the standard Big Bang model \cite{Paradellis-1990,coc2-2004}. Light elements, up-to iron $^{56}$Fe, are belived to be produced by primordial nucleosynthesis\cite{Tytler-2000,kajino-1990,yamamoto-1993,Meibner-2023}. Beyond iron $^{56}$Fe, the elements are produced through various astrophysical processes, including neutron captures, stellar nucleosynthesis, explosive events in supernovas, and many other processes. The nuclei in the proton-rich side of the stability curve are produced by the p-process \cite{rauscher-2012}, and those in the neutron-rich side are produced by the r-process, and the s-process \cite{kappeler-1982}. 
In the study of nucleosynthesis network, the role of a smooth energy-dependent function called astrophysical S-function\cite{alinka-2012,murat-2024} is inevitable since it forms an integral part of the nuclear fusion cross section\cite{Li-2004}. The measurement of the fusion cross-section at astrophysical energies, which lie much below the height of the Coulomb barrier, is a difficult task, as there exists no experimental technique to measure this quantity directly at such low energies. However, there exist limited experimental and analytical techniques to fruitfully explain the experimental findings \cite{Broggini-2010,yakovlev-2010}, and a sound theoretical model is highly needed to explain the experimental findings. Any improvement in the database of fusion cross-section for the light nuclei reaction may give a better picture of nucleosynthesis. By exploiting the advancements in modern computer technology, theoretical calculations of stellar structure and evolution of stellar objects have been performed with greater accuracy. More relevant physical information on stellar objects can, in principle, be extracted by the computational techniques with great precision and soundness.

Motivations for the present work came from several previous studies. Li (2002) explained the fusion cross-sections of the D+T reaction using the Selective Resonant Tunneling Model (SRTM)\cite{Li-2002}. In 2008, Li, Wei, and Liu examined the fusion cross-section of D+T, D+D, and D + $^3$He reactions and compared their results with available 3-parameter, 5-parameter fitting as well as and experimental data \cite{Li-2008}. Later in 2015, Liang, Dong, and Li investigated the fusion cross-section of P+D, P+$^6$Li, P+$^7$Li, D+D, D+T, D+$^3$He, T+T, and T+$^3$He reactions \cite{liang-2015}. Khan {\it et. al.} (2023), were the first to use a space-varying complex potential to explore the fusion cross-section and astrophysical S-factor of some light nuclei, and their findings were compared with the experimental data found in the literature \cite{khan-2023}. The present work is aimed to investigate fusion cross-section and astrophysical S-factor for three important astrophysical reactions:  $^{3}$He ($^3$He,2p)$^{4}$He, $^{6}$Li($^3$He,d)$^{7}$Be and $^{10}$B($^3$He,n)$^{12}$N. It is worth mentioning here that in primordial nucleosynthesis $^3$He($\alpha$,$\gamma$)$^7$Be reaction constitute the main part of the P-P II branch \cite{krane-2010,signore-1999,balbes-1995}. As an improvement over similar earlier works, here a density dependent nucleus-nucleus interaction potential has been invoked in the theoretical computation for all of the three reactions-$^{3}$He($^3$He,2p)$^{4}$He, $^{6}$Li($^3$He,d)$^{7}$Be and $^{10}$B($^3$He,n)$^{12}$N. The paper is organized as follows:

In Section 2, we have introduced the theoretical framework for the astrophysical S-factor and fusion cross-section ($\sigma$) appropriate to the low energy regime. Results and discussions have been presented in Section 3, in which comparisons of theoretical results with corresponding experimental observations have been incorporated. Finally, Section 4 has been closed with a brief summary and conclusions.
	
\section{Theoretical Framework:}
	
A transparent knowledge and clear understanding of the cross-section $\sigma$(E) of the nuclear fusion reaction is important for having the broad picture of the nucleosynthesis network \cite{C. Rolfs 2007}. In nuclear astrophysics aspects, the Coulomb barrier plays a pivotal role \cite{keeley-2007} in the fusion process through sub-barrier tunnelling, which occurs in the stellar energy regime \cite{J. P. Eisenstein 1992}. Though the temperature is not sufficient for a direct fusion reaction, the reaction happens, and radiant energy is produced. This triggered the researchers to believe in the crucial role of the quantum mechanical tunnelling phenomenon\cite{trixler-2013}. 

For the computation of the Gamow penetration factor \cite{T. Spillane 2007}, here, a three-component nucleus-nucleus potential including Coulomb, nuclear, and centrifugal repulsion has been used. Astrophysical S-function, S(E) \cite{M.I. Hussein 2020}, which is a crucial astrophysical observable associated with the fusion cross-section $\sigma$(E), is used to handle the energy dependence of the cross-section. In earlier similar calculations, a single-step Selective Resonant Tunnelling Model (SRTM) was used by Liang {\it et. al.} (2015)\cite{Liang-2015} to compute fusion cross-section for D+T reaction using a complex square well potential. Li {\it et. al.} (2004)\cite{Li-2004} used SRTM to compute D+D and D+$^3$He reaction fusion cross-section. Singh {\it et. al.} (2019)\cite{Singh-2019} calculated fusion cross-section and astrophysical S-factor for D+D, D+T, D+$^3$He using  SRTM. Khan {\it et. al.} (2023)\cite{Khan-2023} also computed $\sigma(E)$ and S(E) for D+D and p+$^{11}$B by the SRTM, introducing a phenomenological space-varying complex potential instead of the constant square-well complex potential used in earlier similar works. 

Durant {\it et. al.} (2018)\cite{durant-2018} by quantum tunnelling model computed fusion cross-section of $^{16}$O + $^{16}$O reaction by employing a microscopically derived density dependent double folding potential model. This density-dependent potential has a smooth surface profile, capable of giving a better fit to experimentally observed scattering and fusion cross-section data. On the other side, the purely phenomenological energy-independent complex square-well potential has a sharp nuclear boundary, constant depth, and it ignores surface diffuseness. Further, the density-dependent double folding potential has better predictive power over the phenomenological complex square-well potential. Apart from the above, it can be utilised to describe the structure of exotic halo nuclei.

Double-folding model potential being density-dependent, knowledge of the density distribution of interacting nuclei is essential. This density distribution may have several forms, like a Fermi distribution, a Gaussian distribution, or Variational Monte Carlo(VMC)\cite{Sert-2021}. For the present work, we picked up  the two-parameter Fermi distribution function, which is also known as the Woods-Saxon distribution. This choice has been made with an eye towards its possible application for the reaction involving heavier isotopes in the later stage. Out of several possible choices for the nucleon-nucleon (NN) interactions commonly adopted in the derivation of double folding potential, like the Michigan-3-Yukawa (M3Y) potential variants M3Y-Reid and M3Y-Paris\cite{chushnyakova-2021}; the São Paulo potential (SPP2)\cite{chamon-2021} and density-dependent potentials CDM3Y6\cite{guo-2013}, we have picked the M3Y-Reid for this work. Though this double-folding model potential is independent of nuclear charge and nuclear spin, it depends on the matter distribution of interacting nuclei. Apart from being capable of describing the mean field interaction between the colliding nuclei, this microscopic double-folding potential also provides a clearer picture of the interaction between the target and projectile nucleus. The basic inputs in the double-folding potential model calculation are the nuclear densities of the colliding nuclei and an appropriate nucleon-nucleon interaction between the nucleons of the projectile and the target. 

\begin{figure}[htb]
	\centerline{\includegraphics[width=0.65\textwidth]{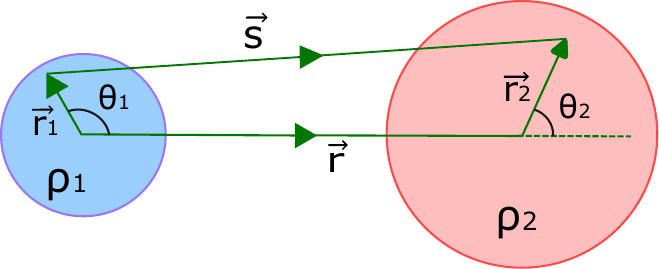}} 
	\caption[]{Coordinates used in double-folding nucleus-nucleus potential of the colliding system.} 
	\label{f1}
\end{figure}

To test the appropriateness of the chosen potential model, few $^3$He-induced fusion reactions, namely- $^{3}$He($^3$He,2p)$^{4}$He, $^{6}$Li($^3$He,d)$^{7}$Be and $^{10}$B($^3$He,n)$^{12}$N are chosen in which $^3$He  projectile is bombarded on targets $^3$He, $^{6}$Li, $^{10}$B, respectively. As already stated in the preceding paragraph, for the investigation of nuclear reactions using double-folding potential, density distributions of the projectile and target nuclei are crucial. In the nucleus-nucleus optical model potential, the double folding potential represents its real part \cite{C.L. Woods 1982}. Such a double-folding potential model, with realistic nucleoin the double-folding potential calculationsn-nucleon interaction based upon a G-matrix constructed from the Reid potential, has been used earlier to calculate the real part of the optical potential for heavy-ion scattering \cite{Satchler-1979}. In the direct channel, the double-folding potential \cite{Satchler-1990} is calculated by integrating the nucleon-nucleon interaction over the $^3_2$He and $^4_2$He density distributions $\rho$$_1$(r$_1$) and $\rho$$_2$(r$_2$) of the colliding nuclei and given by
\begin{equation}\label{Eq:DFP}
	U_{DF}(r)=\int\int{\rho_1}(r_1){\rho_2}(r_2)v_{NN}(s)d^3r_1d^3r_2
\end{equation}
Where $r$ is the distance between the projectile and target nuclei, $r_1$ and $r_2$ are the radial distances of the nucleons of the projectile and target nuclei, respectively, measured from their respective centres. Eq. (\ref{Eq:DFP}) involves an integral over radial and angular variables. The angle between $\vec{r}_1$ and $\vec{r}$ denoted as $\theta_1$, that between $\vec{r}_2$ and $\vec{r}$ is denoted as $\theta_2$ and the angle between $\vec{r}_1$ and $\vec{r_2}$ is $\propto (\theta_1-\theta_2)$ and the vectorial representation of the nucleon-nucleon sepration ($s$), in terms of $r, r_1, r_2, r$ as depicted in Figure \ref{f1} follows: 
\begin{equation}\label{Eq:vec}
	\vec{s} = \vec{r} - \vec{r}_1 +\vec{r}_2 
\end{equation}

The two-parameter Fermi distribution of nuclear matter is given by Eq. (\ref{Eq:rho1})
\begin{equation}\label{Eq:rho1}
	\rho(r) = \frac{\rho_o}{1+\exp(\frac{r-c}{a})}, 
\end{equation}\\
in which $\rho_o$ = 0.1507 (fm$^{-3}$), $c = R_0A_{x}^{1/3}$; \textit{x = t,} for target or \textit{x = p,} for projectile; $a$ = 0.486 (fm).
The DDM3Y-Reid\cite{amer-2016,murat-2024} nucleon-nucleon realistic interaction used in the above double folding calculation (see Eq. (\ref{Eq:DFP})) is represented by 
\begin{equation}\label{Eq:re}
	v_{NN}(r) = v(r) + J_{00}\delta{(r)}
\end{equation} 
where $v(r)$ represents the Yukawa-type effective nucleon-nucleon interaction potential function that consists of a short-range repulsive component and a relatively longer-range attractive component, given by
\begin{equation}
	v(r) = 7999 \frac{\exp(-4r)}{4r} - 2134\frac{\exp(-2.5r)}{2.5r}                           
\end{equation}
The nucleon exchange term $J_{00}$ has linear energy dependence as in Eq. (\ref{Eq:re}) is defined in Eq. (\ref{exc}) as 
\begin{equation}\label{exc}
	J_{00} = -276 [1-\frac{0.005 E_{lab}}{A_p}] \: MeVfm^3
\end{equation}
A complex nucleus-nucleus potential $U_N(r)$, may be expressed as 
\begin{equation}\label{e1}
	U_N(r)= \epsilon_r U_{DF} (r) +i \epsilon_i U_{iDF} (r)  
\end{equation} 
where the imaginary component is derived using the expression
\begin{equation}
	U_{iDF} (r) = - \frac{1}{r} \frac{d}{dr} (U_{DF} (r))  
\end{equation}
For the achievement of the desired match with observed data, we have deliberately introduced two controlling parameters, namely, $\epsilon_r$ and $\epsilon_i$, in the nuclear optical potential represented by Eq. (\ref{e1}). The total nucleus-nucleus potential for the colliding system may be defined as 
\begin{equation}\label{Eq:eff}
	U(r)  = U_{Coul}(r) +U{_{Cf}}(r) + U{_N} (r)
\end{equation}
The first term on the right-hand side of Eq. (\ref{Eq:eff}) denotes the Coulomb potential that depends on the charges of the projectile and target in the colliding system\cite{Satchler-1983} and is represented by
\begin{equation}\label{Eq:Cou} 
	U_{Coul}(r)=
	\begin{cases}
		1.44 \frac{Z_1Z_2}{r};                   \text{for  r $>$ R      [MeV]}    \\
		1.44 \frac{Z_1Z_2}{2R}(3 - \frac{r^2}{R^2});  \text{for r$ \le$ R      [MeV]}                  
	\end{cases}
\end{equation}
In Eq. (\ref{Eq:Cou}), $Z_1$ and $Z_2$ are the charges of the projectile and target particles. Height of the Coulomb barrier for two approaching nuclei of mass numbers $A_1$ and $A_2$ is determined by Eq. (\ref{CB})
\begin{equation}\label{CB}
	V_{CB}=\frac{e^2}{4\pi\epsilon_0}\frac{Z_1Z_2}{R} = (1.44)\left(\frac{Z_1Z_2}{{R_0(A^{1/3}_1+A^{1/3}_2)}}\right)
\end{equation}
where $R_0$ is the nuclear radius parameter and $R$ is the touching distance between the centres of the colliding nuclei. The second term of Eq. (\ref{Eq:eff}) denotes the centrifugal potential\cite{Sert-2021} given by
\begin{equation}\label{Eq:c}
	U{_{Cf}}(r) =  \frac{\hbar^2 l(l+1)}{2 \mu r^2}
\end{equation}
where $l$ is the angular momentum and $r$ is the radial variable.

The Schr\"{o}dinger equation for the total interaction potential $U(r)$ is given by  
\begin{equation}\label{Eq:sc}
	(-\nabla^2+   \frac {2 \mu U(r)}{\hbar^2} - k^2) \psi(r)=0,                                    
\end{equation}
where $k = \sqrt{\frac{2\mu E}{\hbar^2}}$; $E$ is the energy of the relative
motion in the center of mass system; $\mu$ is the reduced mass of the colliding nuclei, $\hbar$ is the reduced Planck’s constant, and $\psi$(r) represents the sum of the nuclear and Coulomb wave function:
\begin{equation}\label{Eq:psi2}
	\psi(r) = \psi_N(r)  + \psi_C(r)                                                   
\end{equation}

In Eq. (\ref{Eq:psi2} ), the second term, i.e., the Coulomb wave function, holds only the incoming wave, although in the asymptotic range the nuclear wave function appears for the outgoing wave. Eq. (\ref{Eq:sc}) has been reduced to an $r$-dependent single differential equation represented by Eq. (\ref{Eq:one}).

\begin{equation}\label{Eq:one}
	( - \frac{d^2}{dr^2}  + \frac{l(l+1)}{r^2}   + \frac{2 \mu U(r)}{\hbar^2}  - k^2) \psi (r) =  0
\end{equation}
When a projectile particle (e.g., $^3$He) is injected into the target nucleus (e.g., $^3$He, $^{6}$Li, $^{10}$B, etc.) to initiate a reaction, the relative motion of the corresponding projectile-target composite system can be described in terms of the wave function $\psi$(r). And the general solution $\psi(r, t)$ of the corresponding Schr\"{o}dinger Eq. (\ref{Eq:one}) for the interacting nuclei may be represented as Eq. (\ref{Eq:si}).

\begin{equation}\label{Eq:si}
	\psi(r, t) = {\frac{1}{\sqrt{4\pi r}} \psi(r) \exp(- \frac{iEt}{\hbar})}                                    
\end{equation}

The fusion cross section ($\sigma$), for projectile energy ($E$) much below the Coulomb barrier, where the classical turning point is much larger than the nuclear radius, barrier penetration probability can be approximated by $\exp(-2\pi\zeta)$ 
so that the charge-induced fusion cross-section can be given by Eq.  (\ref{Eq:sigma}).

\begin{equation}\label{Eq:sigma}
	\sigma(E)=\frac{S(E)} {E} \exp(-2\pi\zeta)  
\end{equation}
where $S(E)$ is the astrophysical S-function and $\zeta$ is the Sommerfeld parameter defined as

\begin{equation}\label{Eq:zeta}
	\zeta=\frac{Z_1Z_2e^2} {\hbar v}  
\end{equation}

In SRTM approach which employs a complex nuclear potential including Coulomb and centrifugal terms, the reaction cross-section ($\sigma$) can be calculated in terms of the phase shift, $\delta_0$, introduced by the nuclear potential in the wave function at the low energy limit (where only the S-wave contributes) using the Eq. (\ref{Eq:psi3}) 

\begin{equation}\label{Eq:psi3}
	\sigma = \frac{\pi}{k^2} (1-|\eta|^2)                                                   
\end{equation}
where $\eta$ = e$^{2i\delta_0}$ and $k$ is the wave number corresponding to the relative motion. Since the nuclear potential is a complex one, the corresponding phase shift $\delta_0$ is also a complex number and can be indicated as Eq. (\ref{cot})
\begin{equation}\label{cot}
	\cot(\delta_0) = W_r + iW_i                                                 
\end{equation}
where $W_r$ and $W_i$ are connected to the real and imaginary components of the complex wavenumber corresponding to the complex nuclear potential.
The wavenumber corresponding to the complex nuclear potential can be expressed as
\begin{equation}
	\begin{split}  
		K(r)&= \sqrt{\frac{2\mu}{\hbar^2}[U(r)-E]}\\
		&= \sqrt{\frac{2\mu}{\hbar^2}[U_r(r)-E+iU_i(r)]}\\
		%   &=\sqrt{\frac{2\mu}{\hbar^2}[U_C(r)+U_{Cent}(r)+U_{N}(r)-E]}\\
		%   &=\sqrt{\frac{2\mu}{\hbar^2}[U_C(r)+C_rU_{DF}(r) +iC_iU_{iDF}(r)-E]};[\textit{for l=0}]\\
		&=\sqrt{\frac{2\mu}{\hbar^2}[U_r(r)-E]}[1+i\frac{U_i(r)}{U_r(r)-E}]^{1/2}\\
		&\simeq \sqrt{\frac{2\mu}{\hbar^2}[U_r(r)-E]}+\frac{i}{2} \sqrt{\frac{2\mu}{\hbar^2}}\frac{U_{i}(r)}{\sqrt{U_r(r)-E}}\\ 
		&=K_r(r)+iK_i(r)
	\end{split}                                                  
\end{equation}
$U_r(r)$=Real Part of $U(r) = U_{Coul}(r)+U_{Cf}(r)+\epsilon_rU_{DF}(r)$ and $U_i(r)$=Imaginary Part of $U(r) = \epsilon_iU_{iDF}(r)$. At the matching radius $r=R$, $KR = K_rR+iK_iR$ and it may be represented as $Z = Z_r + iZ_i$ for $Z = KR$, $Z_r = K_rR$, $Z_i = K_iR$. \\
In terms of real and imaginary parts of $Z$, the pair of parameters $W_r$ and $W_i$ for the $l=0$ partial wave may be designated as
\begin{equation}
	\begin{split}
		W_r&=\Omega^2 [\frac{R_c}{R} \frac{Z_r \sin(2 Z_r) + Z_i \sinh(2Z_i)}{2[\sin^2(Z_r)+\sinh^2(Z_i)]}] 
		-2\Omega^2[ln(\frac{2R}{R_c}) +2C +h(kR_c)]\\
		W_i&= {\Omega^2 [\frac{R_c}{R} \frac{Z_i \sin(2 Z_r) - Z_r \sinh(2Z_i)}{2[\sin^2(Z_r)+\sinh^2(Z_i)]}]}  
	\end{split} 
\end{equation}
where $R_c = \frac{\hbar^2}{Z_1Z_2\mu e^2}$ is the Coulomb unit of length and $C = 0.577$ is Euler’s
constant. The energy dependent function $h(kR_c)$ is connected to the logarithmic derivative of $\Gamma$ function: \cite{Li-2002, Li-2004}
$$h(kR_c)\equiv Re[\psi(\frac{-i}{kR_c})]+\log(kR_c);\: \psi(x)\equiv\frac{\Gamma^{\prime}(x)}{\Gamma(x)}$$ where [$\Gamma(x)$ is the Gamma function] and\\
\begin{equation}
	h(y)=\frac{1}{y^2} \sum_{s=1, 2, .., }^{\infty}\frac{1}{s(s^2+y^{-2})} -C +ln(y)
\end{equation}

The fusion cross-section can then be expressed as

\begin{align}\label{b}
	\sigma& = \frac{\pi}{k^2} \left\{\frac{-4W_i}{W_r^2+(W_i-1)^2}\right\} \nonumber\\
	&= {\frac{\pi}{k^2}\frac{1}{\Omega^2} \left\{\frac{-4\omega_i}{\omega_r^2+(\omega_i-\frac{1}{\Omega^2})^2}\right\}}  
\end{align}
where the quantity $\Omega^2$=$\frac{\exp(\frac{2\pi}{kR_c})-1}{2\pi}$ is related to the Gamow penetration factor.

The last factor in Eq. (\ref{b}) within curly braces \{\} is called the astrophysical S-factor ($S(E)$), which is a smooth, slowly varying function of energy ($E$) given by Eq. (\ref{sf})
\begin{equation}\label{sf}
	S(E) = \frac{-4\omega_i}{\omega_r^2+(\omega_i-\frac{1}{\Omega^2})^2}
\end{equation}
$S(E)$ facilitates its extrapolation down to astrophysical energies. In the above Eqs. (\ref{b}) \& (\ref{sf}), $\omega = \omega_r + i\omega_i$ = $W/\Omega^2$ = ($W_r+iW_i)/\Omega^2$. The wave function inside the nuclear well (ie, in the region $r \le R$) is controlled via the parameters- $c_r$ and $c_i$ introduced in Eq. (\ref{e1}). The Coulomb wave function outside the nuclear well ($r \ge R$) is determined by two other parameters: the real and the imaginary part of the complex phase shift $\delta_{0r}$ and $\delta_{0i}$. A pair of convenient parameters, $W_r$ and $W_i$, is introduced
to make a linkage between the cross section and the nuclear potential. This facilitates
a clear understanding of the resonance and the selectivity in damping.
The continuity of the wave function at the boundary ($r = R$) can be expressed
by the matching of the logarithmic derivative of the wave function. In the above
relations, $k$ represents the energy-dependent wavenumber outside the nuclear well, and $R_c$ is the constant Coulomb unit of length. Motivated by the work of Canon {\it et. al.} (2002)\cite{canon-2002}, an exponential fit to the computed astrophysical S-factor data is done, and the corresponding fitting formula is given by Eq. (\ref{eq-26}) 
\begin{gather}\label{eq-26} 
	S(E)=S_0 + S_1 \exp(-E/S_2)
\end{gather}
where $E$ is the center of mass energy, and $S_{\nu} (\nu = 0,1,2)$ are the fitting parameters.

\section{Results and Discussions: }
Thermonuclear fusion reaction is considered to occur through quantum mechanical tunneling of the interacting particles through the Coulomb barrier \cite{rubakov-1984} appearing due to their charges when the energy is lower than the barrier peak value. An important observable relevant to the investigation of nuclear fusion reactions in the sub-barrier energy regime is the fusion cross-section that has significant implications in astrophysics \cite{E. Yildiz 2020}. Computation of fusion cross-section at sub-barrier energies involves another important astrophysical observable known as the astrophysical S-function. A bright star has its brightness due to the power generated in the belly of the star at the cost of nuclear fuel (mainly protons) that are used in the nuclear fusion reaction. The proton-proton chain reactions are the primary processes that produce most of the energy in low-mass stars \cite{bethe-1939,Salpeter 1952}. At the starting phase, two protons fused to form a deuterium nucleus, releasing a positron and an electron neutrino as by-products. In the next stage, the produced deuterium captures a proton to produce a helium-3 nucleus and an energetic gamma photon. Later, two interacting helium-3 nuclei fused together to produce a helium-4 nucleus (known as an n $\alpha$-particle), ejecting two protons. In subsequent events, a helium-3 nucleus and a helium-4 nucleus undergo fusion to yield beryllium-7 and a gamma photon \cite{E. G. Adelberger et al 2011}. There are several other light-nuclear fusion reactions involved in nucleosynthesis and energy production in stars. For the present study we have chosen three $^3$He-induced fusion reactions- $^{3}$He($^3$He,2p)$^{4}$He, $^{6}$Li($^3$He,d)$^{7}$Be and $^{10}$B($^3$He,n)$^{12}$N. For the computation of the fusion cross-section $\sigma(E)$ and astrophysical S-function, we have adopted the formalism of the selective resonant tunneling model approach\cite{khan-2023} using the density-dependent nucleus-nucleus potential obtained by double-folding calculation. In the double folding calculation of the nucleus-nucleus potential, we have made use of DDM3Y-Reid NN potential\cite{chushnyakova-2021} together with the Woods-Saxon density distribution. A few adjustable parameters, namely the values of the real and imaginary components of the effective nucleus-nucleus potential at the matching boundary ($r=R$) and the nuclear size parameter $R_0$, appear in the theoretical computation of the fusion cross-section and astrophysical S-function. 
For all three reactions being investigated in the present work, the values of the real and imaginary components of the effective potential at the matching radius ($r=R$) have been presented in columns 2 \& 3 of Table \ref{tab-01}. 
As the nuclear size parameter $R_0$ affects the complex phase shift $\delta_0$ defined in Eq. (\ref{cot}), which in turn affects the fusion cross-section $\sigma$ given in Eq. (\ref{Eq:psi3}) and astrophysical S-function $S(E)$, we have adjusted value of $R_0$ to get the desired result. Representative values of adopted $R_0$ values are also listed in column 4 of Table \ref{tab-01} for the reactions studied. The astrophysical S-factor can be computed using Eq. (\ref{sf}). It is to be noted that the fusion cross-section ($\sigma$) can be computed in two ways: i) directly following Eq. (\ref{b}), and ii) following Eq. (\ref{Eq:sigma}) via use of $S(E)$ obtained by an exponential fitting of computed data. The corresponding fitting formula is given by Eq. (\ref{eq-26}) and the fitting parameters for individual reactions are listed in columns 5 to 7 of Table \ref{tab-01}. 

\begin{table}[htbp]
	\caption[]{Values of the parameters used for the calculation of cross-section $\sigma(E)$ and astrophysical S-factor $S(E)$.}\small\smallskip
	\tabcolsep=3.6pt
	\begin{tabular}{@{}ccccccc@{}}
		\hline
		\hline
		&&&\\[-8pt]
		\textbf{Nuclear}&\textbf{U$_{DF}(R)$}&\textbf{U$_{iDF}(R)$} &\textbf{R$_0$} &\textbf{S$_0$}&\textbf{S$_1$}&\textbf{S$_2$}\\
		\textbf{reactions}&\textbf{(MeV)}&\textbf{(keV)}  &\textbf{(fm)}&\textbf{(keV-mb)}&\textbf{($\mu$b)}&\textbf{($\mu$b/keV)} \\
		\hline
		&&\\[-8pt]$^{3}$He($^3$He,2p)$^{4}$He& -156.48 & -641.45 &0.80   & 2.325 & $4.6\times 10^{16}$ & $0.81$\\
		$^{6}$Li($^3$He,d)$^{7}$Be& -34.92 & -410.50 &1.71 & 86.008 & $1111.95 $& $598.98$ \\
		$^{10}$B($^3$He,n)$^{12}$N & -40.95 & -731.27 &1.36      & 18.894&  $40.54$ & $2035.91$ \\
		\hline
		\hline
	\end{tabular}
	\label{tab-01}
\end{table}

\begin{figure}[htb]
	\centerline{\includegraphics[width=0.65\textwidth]{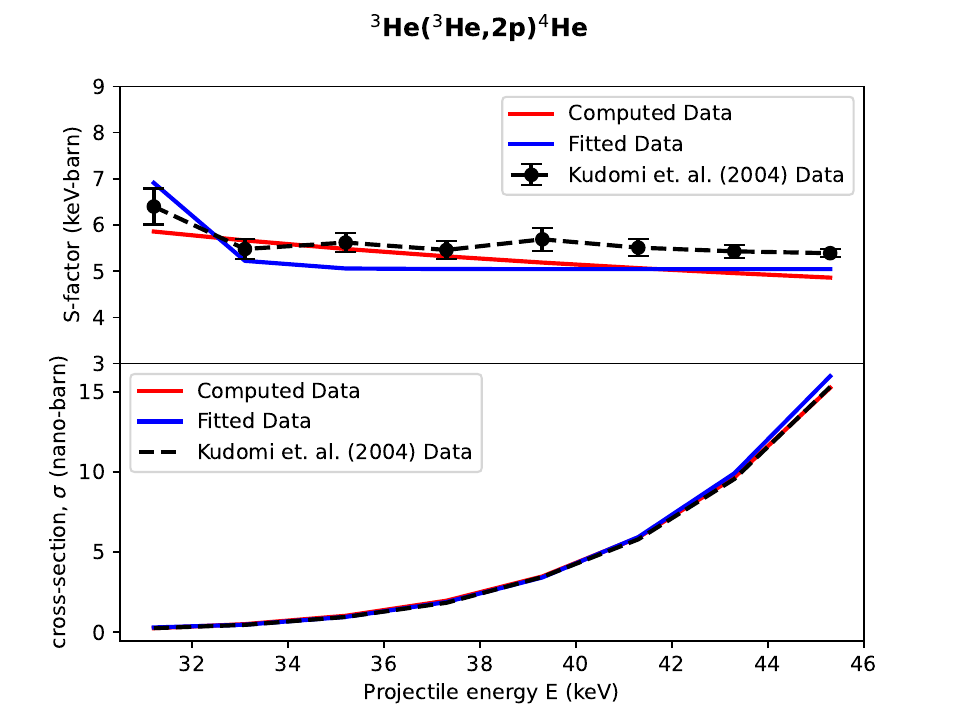}} 
	\caption[]{Upper panel: Comparison of computed S-factor data with experimentally observed data reported by Kudomi et. al. (2004)\cite{kudomi-2004} Lower panel: Comparison of computed fusion cross-section data with experimentally measured data reported by Kudomi et. al. (2004)\cite{kudomi-2004} for $^{3}$He($^3$He,2p)$^{4}$He reaction.}
	\label{fig2}
\end{figure}

The $^{3}$He($^3$He,2p)$^{4}$He nuclear reaction plays a significant role in the pp-II and pp-III chains and dominates the power production in low-mass stars at the core temperatures around 10-20 million Kelvin. This nuclear reaction also acts as a key player in the building up of $^4$He in the stellar core and affects star lifetimes and evolution\cite{Dwarakanath-1971, Dwarakanath-1974}. The production rates of $^4$He in stellar cores confirm the quantum tunneling \cite{rubakov-1984} in nuclei and weak interactions. We have calculated the astrophysical S-factor and fusion cross-section at sub-barrier energies of the reaction $^{3}$He($^3$He,2p)$^{4}$He, and the representative set of computed S-factor and cross-section data (obtained for $\epsilon_r=4.5, \epsilon_i=0.02$) are respectively presented in columns 4 \& 6 of Table \ref{tab-02}. The variation of S-factor with projectile energy is depicted in the upper panel of Figure \ref{fig2}. The experimentally observed S-factor data reported by Kudomi {\it et. al.} (2004)\cite{kudomi-2004} are also plotted with a broken line in the upper panel of Figure \ref{fig2} and also presented in column 3 of Table \ref{tab-02}. Variation of the fusion cross-section ($\sigma(E)$) versus energy ($E$) is shown in the lower panel of Figure \ref{fig2}. The observed data reported by Kudomi {\it et. al.} (2004)\cite{kudomi-2004} are also plotted with a broken line in the lower panel of Figure \ref{fig2} and also shown in column 5 of Table \ref{tab-02}. Comparison of the experimentally observed S-factor data presented in column 3 with the computed S-factor data presented in column 4 of Table \ref{tab-02} indicates a good agreement between. The same trend of agreement can also be seen by a comparison of the graphs shown in the upper panel of Figure \ref{fig2}. Comparison of the experimentally observed cross-section data of Kudomi {\it et. al.} (2004)\cite{kudomi-2004} presented in column 5 of Table \ref{tab-02} with the computed data presented in columns 6 \& 7 of Table \ref{tab-02} exhibits a nice agreement among themselves. The same pattern of agreement can also be seen from the graphs shown in the lower panel of Figure \ref{fig2}. 

\begin{table}[htbp] 
	\caption[]{S-factor ($S(E)$) and cross-section ($\sigma(E)$) data of  $^{3}$He($^3$He,2p)$^{4}$He fusion reaction}\small\smallskip 
	\tabcolsep=3.6pt
	%\arrayrulecolor{red}{ 
		\begin{tabular}{@{}ccccccrc@{}}
			\hline
			\hline
			&&&\\[-8pt]
			\textbf{Nuclear}&\textbf{Energy}&\textbf{$S$(keV-mb)}&\textbf{$S$(keV-mb)}&\textbf{$\sigma$(nb)} &\textbf{$\sigma$(nb)}&\textbf{$\sigma$(nb)}  \\
			\textbf{reaction}&\textbf{(keV)}&\textbf{Expt.\cite{kudomi-2004}}&\textbf{Eq.(\ref{sf})}&\textbf{Expt.\cite{kudomi-2004}} &\textbf{Eq. (\ref{b})}&\textbf{Eq. (\ref{Eq:sigma})} \\ 
			\hline
			&&&\\[-8pt]$^{3}$He+$^3$He 
			& 31.2 &6.40 $\pm$ 0.39&5.86& 0.246 & 0.245&0.291 \\    
			& 33.1 &5.48 $\pm$ 0.22&5.67&0.452 &0.498&0.463\\
			& 35.2 &5.62 $\pm$ 0.21&5.48& 0.964 &1.019&0.949\\ 
			&37.3 &5.46 $\pm$ 0.20&5.32& 1.831 &1.960&1.876 \\
			& 39.3 &5.69 $\pm$ 0.25&5.19& 3.440&3.475&3.412 \\
			&41.3 &5.51 $\pm$ 0.18&5.07& 5.790 &5.901&5.931\\
			& 43.3 &5.43 $\pm$ 0.14&4.96& 9.550 & 9.648&9.907\\ 
			& 45.3 &5.39 $\pm$ 0.09&4.86& 15.300 &15.248&15.973\\      
			\hline
			\hline
		\end{tabular}
		\label{tab-02} 
	\end{table}
	
\begin{figure}[htb]
		\centerline{\includegraphics[width=0.65\textwidth]{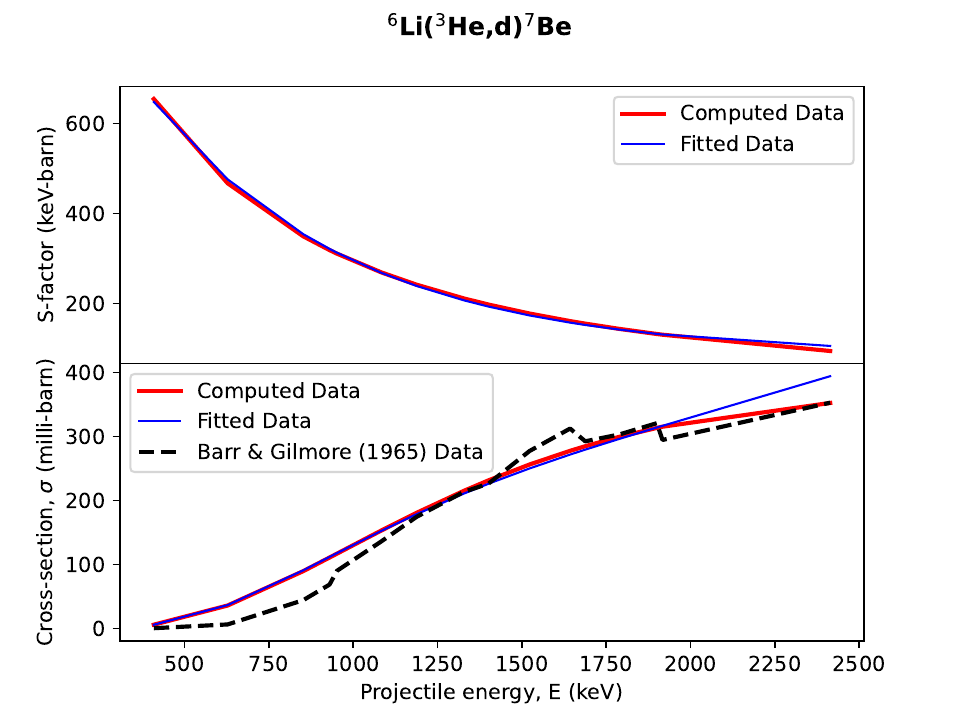}} 
		\caption[]{Upper panel: Astrophysical S-factor computed by using SRTM formula (Eq.(\ref{sf})). Lower panel: Comparison of computed fusion cross-section data with experimentally measured data reported by Barr and Gilmore (1965)\cite{barr-1965} for $^{6}$Li($^3$He,d)$^{7}$Be reaction.}
		\label{fig3}
	\end{figure}

	The fusion reaction $^{6}$Li($^3$He,d)$^{7}$Be is also an important one that played a crucial role in the Big Bang nucleosynthesis (BBN). Though BBN predicts lithium abundances, its scientific analysis reveals a significant discrepancy (termed as "lithium problem") where primordial $^7$Li is less abundant compared to stellar abundances. The rate of $^6$Li($^3$He,d)$^7$Be reaction helps constrain how much $^6$Li is used up, refining BBN measurements and calculations. The $^6$Li($^3$He,d)$^7$Be reaction plays a pivotal role in the understanding of the cosmological puzzle arising from lithium isotope evolution in the universe\cite{Ludecke-1968,Bertulani-2019}. Representative values of the computed astrophysical S-factor and fusion cross-section ($\sigma$) at sub-barrier energies of the reaction $^6$Li($^3$He,d)$^7$Be (obtained for $\epsilon_r=1.48, \epsilon_i=0.01$) are respectively presented in columns 3 to 4 \& 6 to 7 of Table \ref{tab-03}. The pattern of variation of S-factor with projectile energy is shown in the upper panel of Figure \ref{fig3}. The S-factor data computed by using Eq. (\ref{sf}) is shown with a red color line while those obtained by Eq. (\ref{eq-26}) are shown with a blue color line in the upper panel of Figure \ref{fig3}. The corresponding S-factor data are also presented in columns 3 \& 4 of Table \ref{tab-03}. The graphical pattern of variation of the fusion cross-section ($\sigma(E)$) with projectile energy ($E$) is shown in the lower panel of Figure \ref{fig3}. The experimentally observed data set of Barr and Gilmore (1965)\cite{barr-1965} is plotted with a broken line in the lower panel of Figure \ref{fig3}, and corresponding data are presented in column 5 of Table \ref{tab-03}. The computed cross-section data, which are obtained following Eq. (\ref{b}) and Eq. (\ref{Eq:sigma}) are respectively presented in columns 6 and 7 of Table \ref{tab-03}. Comparison of the experimentally observed cross-section data of Barr and Gilmore (1965)\cite{barr-1965} presented in column 5 of Table \ref{tab-03} with the computed data presented in columns 6 \& 7 of Table \ref{tab-03} indicates fair agreement among themselves. The same pattern of agreement can also be visualized from the graphs shown in the bottom panel of Figure \ref{fig3}. 
	
	\begin{table}[htbp] 
		\caption[]{S-factor ($S(E)$) and cross-section ($\sigma(E)$) data of  $^{6}$Li($^3$He,d)$^{7}$Be fusion reaction}\small\smallskip 
		\tabcolsep=3.6pt
		%\arrayrulecolor{red}{ 
			\begin{tabular}{@{}cccccrrc@{}}
				\hline
				\hline
				&&&\\[-8pt]
				\textbf{Nuclear}&\textbf{Energy}&\textbf{$S$(keV-mb)}&\textbf{$S$(keV-mb)}&\textbf{$\sigma$(mb)} &\textbf{$\sigma$(mb)}&\textbf{$\sigma$(mb)}  \\
				\textbf{reactions}&\textbf{(keV)}&\textbf{Eq.(\ref{sf})}&\textbf{Eq.(\ref{eq-26})}&\textbf{Expt.\cite{barr-1965}} &\textbf{Eq. (\ref{b})}&\textbf{Eq. (\ref{Eq:sigma})} \\ 
				\hline
				&&&\\[-8pt]$^{3}$He+$^6$Li 
				& 409&654.3 &647.7& 00.62 & 06.08& 06.02 \\   
				& 628 &467.0&475.7& 06.50 & 36.09& 36.75\\
				& 852 &349.2&354.1& 44.42 &89.75& 90.99\\
				& 930 &318.6&321.4& 68.77 & 111.00&111.95 \\
				& 952 &310.7&312.9&  90.28 & 117.07&117.88 \\
				& 1084 &269.0&268.0& 136.40 & 153.33& 152.77\\
				& 1188 &241.8&239.0& 174.50 & 180.82& 178.72\\ 
				& 1328 &211.3&207.1& 213.50 & 215.11& 210.82\\
				& 1401 &197.7&193.2& 226.40& 231.47& 226.23\\ 
				& 1523 &177.7&173.4& 277.40& 256.31& 250.14\\
				& 1643 &160.9&157.6& 312.40& 277.62& 271.78\\  
				& 1688 &155.3&152.4& 292.50& 284.83& 279.50\\ 
				& 1789 &143.6&142.1& 303.20 & 299.52& 296.27 \\
				& 1900 &132.3&132.6& 320.80&313.40&314.01 \\
				& 1917 &130.7&131.3& 294.60 &315.33&316.68 \\
				& 2414 &094.5&105.7& 352.80 &352.41& 394.19 \\ 
				\hline
				\hline
				
			\end{tabular}
			\label{tab-03} 
		\end{table}
		
\begin{figure}[htb]
			\centerline{\includegraphics[width=0.75\textwidth]{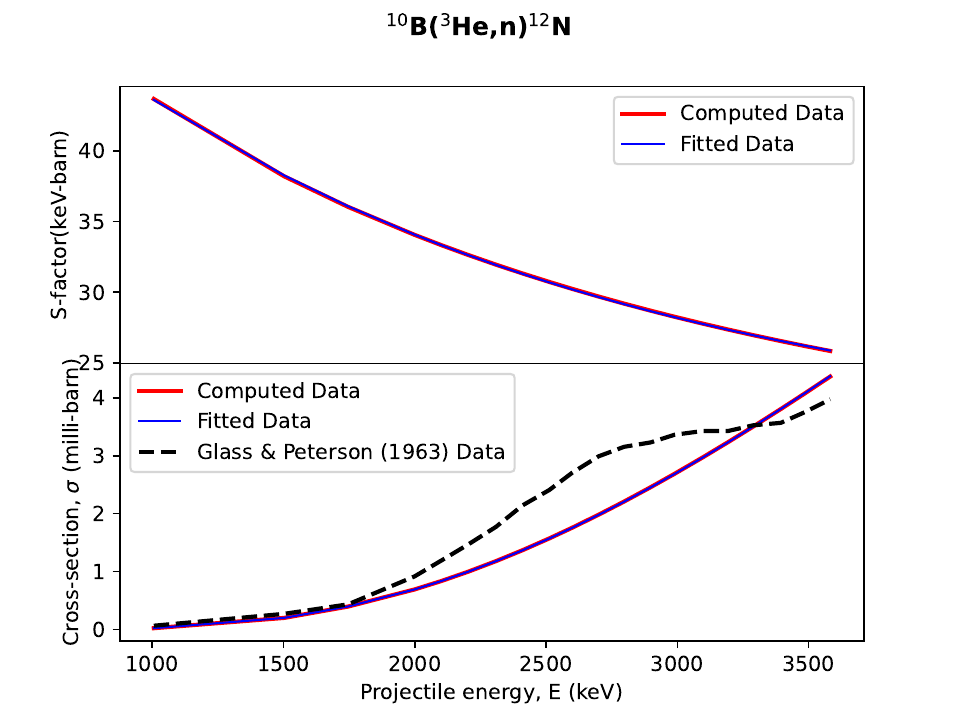}} 
			\caption[]{Upper panel: Astrophysical S-factor computed by using SRTM formula (Eq.(\ref{sf})). Lower panel:Comparison of computed fusion cross-section data with experimentally measured data reported by Glass and Peterson (1963)\cite{glass-1963} for $^{10}$B($^3$He,n)$^{12}$N reaction.} 
			\label{fig4}
		\end{figure}

		Though the third reaction $^{10}$B($^3$He,n)$^{12}$N does not participate in the direct frying pan cooking of elements through nucleosynthesis reactions, it has important implication in the laboratory for providing crucial nuclear physics inputs like reaction rates and nuclear structure information\cite{Fuchs-1974} needed for an accurate modeling of stellar evolution and element production through violent stellar events. It helps explain abundance patterns of s-process elements in stars and Galaxies. Spectroscopic observational analysis of AGB stars indicates enhancements in heavy elements like technetium, which need neutron sources. Models incorporating this reaction may give a better match to the observed s-process yields, particularly in low-metallicity environments where $^{13}$C might be scarce\cite{Zafiratos-1966}. A representative set of computed astrophysical S-factor and fusion cross-section at sub-barrier energies of the reaction $^{10}$B($^3$He,n)$^{12}$N (obtained for $\epsilon_r=1.2, \epsilon_i=0.03$) has been presented respectively in columns 3 to 4 and 6 to 7 of Table \ref{tab-04}. The graphical variation of S-factor with projectile energy is demonstrated in the upper panel of Figure \ref{fig4}. Plots of the fusion cross-section ($\sigma(E)$) versus energy ($E$) are displayed in the lower panel of Figure \ref{fig4}. A set of experimentally observed cross-section data reported by Glass and Peterson (1963)\cite{glass-1963} has been presented in column 5 of Table \ref{tab-04} and has also been plotted with a broken line in the lower panel of Figure \ref{fig4}. As evident from the comparative plots shown in the lower panel of Figure \ref{fig4}, and comparison of data presented in columns 5 to 7 of Table \ref{tab-04}, there is fair agreement among the present computed data and those observed experimentally by Glass and Peterson (1963)\cite{glass-1963}. 
		
		For close comparison we may pick a representative energy and cross-section data (39.3 keV, 3.44 nb) reported by Kudomi {\it et. al.} (2004)\cite{kudomi-2004} against our computed data (39.3 keV, 3.48 nb) and (39.3 keV, 3.41 nb) respectively presented in row 5 of Table \ref{tab-02} for $^{3}$He($^3$He,2p)$^{4}$He reaction. For the fusion reaction $^{6}$Li($^3$He,d)$^{7}$Be we may refer the data (1.328 MeV, 213.5 mb) reported by  Barr \& Gilmore (1965)\cite{barr-1965} against our computed data (1.328 MeV, 215.1 mb) and (1.328 MeV, 210.8 mb) respectively presented in row 8 of Table \ref{tab-03}. And for the reaction $^{10}$B($^3$He,n)$^{12}$N we may refer the data (3.29 MeV, 3.52 mb) reported by  Glass \& Peterson (1963)\cite{glass-1963} against our computed data (3.29 MeV, 3.50 mb) and (3.29 MeV, 3.50 mb) respectively presented in row 4 from the bottom of Table \ref{tab-04}. All of the computed data sets and the experimentally measured data sets are found to be in good agreement with one another.

		\begin{table}[htbp] 
			\caption[]{S-factor ($S(E)$) and cross-section ($\sigma(E)$) data of   $^{10}$B($^3$He,n)$^{12}$N fusion reaction}\small\smallskip 
			\tabcolsep=3.6pt %\arrayrulecolor{red}{ 
				\begin{tabular}{@{}ccccccc@{}} \hline \hline
					&&&\\[-8pt]
					\textbf{Nuclear}&\textbf{Energy}&\textbf{$S$(keV-mb)}&\textbf{$S$(keV-mb)}&\textbf{$\sigma$(mb)} &\textbf{$\sigma$(mb)}&\textbf{$\sigma$(mb)}  \\
					\textbf{reactions}&\textbf{(keV)}&\textbf{ Eq.(\ref{sf})}&\textbf{ Eq.(\ref{eq-26})}&\textbf{Expt.\cite{glass-1963}} &\textbf{Eq. (\ref{b})}&\textbf{Eq. (\ref{Eq:sigma})} \\  
					\hline
					&&&\\[-8pt]$^{3}$He+$^{10}$B
					& 1006&43.69 &43.62&0.065 & 0.022&0.022 \\
					& 1502&38.21 &38.28&0.270 &0.199&0.200\\
					& 1746 &36.04&36.09&0.434 &0.395&0.396\\
					&1999 &34.06&34.08& 0.916 &0.690&0.691\\
					&2101 &33.33&33.33& 1.190 &0.836&0.836\\
					&2202&32.64 &32.64&1.464 &0.995& 0.995\\   
					&2309 &31.95&31.93&1.771 & 1.179& 1.179\\   
					&2411 &31.32&31.29&2.144 &1.369&1.369\\    
					&2512 &30.72&30.69& 2.407&1.571& 1.570\\   
					&2599 &30.23&30.20& 2.704&1.755& 1.754\\   
					&2700 &29.69&29.65& 2.989&1.980& 1.978\\   
					&2797 &29.18&29.15& 3.153&2.20& 2.205\\   
					& 2898 &28.69&28.66& 3.229&2.455&2.452\\   
					& 2989 &28.25&28.23& 3.361&2.686&2.684\\   
					& 3096&27.77&27.75& 3.426&2.968&2.967\\  
					&3192&27.35&27.34& 3.425&3.230&3.229\\
					&3288&26.95&26.95& 3.524&3.499& 3.500\\
					&3394&26.53&26.54& 3.567&3.805&3.807\\
					&3486&26.18&26.21& 3.753&4.076&4.081\\
					&3582&25.83&25.89&3.973&4.366&4.373\\\hline
					\hline
					
				\end{tabular}
				\label{tab-04} 
			\end{table}

\section{Summary and conclusions: }

The main goal set for the present work was to derive a density-dependent complex potential and check its suitability for the computation of astrophysical S-factor and fusion cross-section in the sub-barrier energy regime for a few $^3$He-induced fusion reactions. These reactions are important in nuclear astrophysics studies. For this purpose, we adopted the selective resonant tunneling model (SRTM) following the works of Li {\it et. al.} (2000)\cite{Li-2000}, Liang {\it et. al.} (2015) \cite{Liang-2015}. We are motivated by the SRTM model because of the fact that the well-known compound nucleus model \cite{bohr-1936,ghoshal-1950,sharma-2024} fails to successfully describe the process of fusion of light nuclei at very low energies, since the fusing nuclei may still remember the  phase factor of the wave function describing the system. In the compound nucleus model, the reaction is assumed to proceed in two steps: first, fusing to form the compound nucleus, followed by its decay. Here, we consider the selective resonant tunneling model (SRTM) according to which the tunneling probability itself depends upon the decay lifetime; hence, it is a single-step process. The agreement with the experimental data for the deep sub-barrier fusion of light nuclei also suggests that the tunneling proceeds in a single step.

To accomplish the goals set for the present studies, we have chosen a density-dependent double-folding nucleus–nucleus potential model, which has several important applications in reaction dynamics, nuclear structure studies, astrophysical reaction rate estimations, and microscopic optical model analyses. As the double-folding potential model takes into account nucleon–nucleon (NN) interaction as well as the density distributions of the projectile and the target nuclei, it has an advantage over other available interaction models in the analysis of nuclear structure and in-medium effects in nuclear reactions. Real part of the density-dependent double-folding optical potential governs the barrier shape and quasi-bound state energies inside the potential trap created by the attractive nuclear potential and the repulsive Coulomb barrier, while the imaginary part takes care of the absorption of $^3$He flux causing fusion. Derivation of the double folding potential involves integration on the matter density distribution of the two colliding nuclei and the nucleon-nucleon interaction potential. We have chosen the matter density distribution of the target and projectile, having the nature of a Fermi distribution function. We have picked up the reactions $^{3}$He($^3$He,2p)$^{4}$He, $^{6}$Li($^3$He,d)$^{7}$Be and $^{10}$B($^3$He,n)$^{12}$N for the present study taking note of their remarkable role in nuclear astrophysics. We have computed the astrophysical S-factor and fusion cross-section of $^{3}$He($^3$He,2p)$^{4}$He, $^{6}$Li($^3$He,d)$^{7}$Be and $^{10}$B($^3$He,n)$^{12}$N reactions using density dependent double-folding potential. Both of the projectile (eg $^3$He) and targets (eg $^3$He, $^{6}$Li, $^{10}$B) are assumed to be in their ground states. It can be seen that the present method of studies successfully reproduces the experimental results as evident from the comparative Figures - \ref{fig2}, \ref{fig3}, \ref{fig4} respectively. The same can also checked through a close look at the data presented in columns 5 to 7 of Tables \ref{tab-02} to \ref{tab-04} for the reactions $^{3}$He($^3$He,2p)$^{4}$He, $^{6}$Li($^3$He,d)$^{7}$Be and $^{10}$B($^3$He,n)$^{12}$N respectively. 

Since all of the three $^3$He-induced fusion reactions viz- $^{3}$He($^3$He,2p)$^{4}$He, $^{6}$Li($^3$He, d)$^{7}$Be and $^{10}$B($^3$He,n)$^{12}$N have been analyzed in the sub-barrier energies and the computed data agrees fairly with the corresponding experimentally observed values, we may say that the preliminary goals set for the present studies have been accomplished. The microscopic density-dependent double-folding potential developed here will open the gateway for further advancements towards the investigation of reactions involving heavier nuclei, as well as for model refinements for more complex non-resonant and resonant reactions. 

\textbf{Acknowledgements}\\
The authors are thankful to Aliah University for providing the computational facilities. One of the authors, Md. A. Khan, acknowledges a research grant from Ausandhan National Research Foundation (ANRF) (Erstwhile Department of Science and Technology (DST)), Govt. of India under the State University Research Excellence (SURE) Scheme, having grant No. SUR/2022/004670.

\textbf{Competing interests:}
The authors hereby declare that no competing interests exist.

\end{document}